\documentclass[floatfix,twocolumn,showpacs,preprintnumbers,amsmath,amssymb,prb,superscriptaddress]{revtex4}
\newcommand{\etal}{{\it et al. }}
\usepackage{graphicx}
\usepackage{epstopdf}
\usepackage{dcolumn}

\begin{document}

\preprint{Draft version \today}

\title{The superfluid-insulator transition in the disordered two-dimensional Bose-Hubbard
model}

\author{Fei Lin \footnote{Current address: Department of Physics, Robeson Hall, Virginia Tech,
Blacksburg, VA 24061-0435, USA}}
\affiliation{Department of Physics, University of Illinois at Urbana-Champaign, Urbana, IL 61801, USA}
\author{Erik S. S{\o}rensen}
\affiliation{Department of Physics and Astronomy, McMaster University, Hamilton, Ontario,
 Canada L8S 4M1}
\author{D. M. Ceperley}
\affiliation{Department of Physics, University of Illinois at
Urbana-Champaign, Urbana, IL 61801, USA}

\date{\today}
	
\begin{abstract}
We investigate the superfluid-insulator transition in the
disordered two-dimensional Bose-Hubbard model through
quantum Monte Carlo simulations. The Bose-Hubbard model is
studied in the presence of site disorder and the quantum
critical point between the Bose-glass and superfluid is
determined in the grand canonical ensemble at $\mu/U=0$ (close to $\rho=0.5$), $\mu/U=0.375$ (close to
$\rho=1$), and $\mu/U=1$ as well as in the  canonical ensemble at $\rho=0.5$ and $1$.
Particular attention is paid to disorder averaging and it is
shown that a large number of disorder realizations
are needed in order to obtain reliable results. Typically,
more than $100,000$ disorder realizations were used. In the
grand canonical ensemble, we find $Z t_c/U=0.112(1)$ with
$\mu/U=0.375$, significantly different from previous studies.
When compared to the critical point in the absence of disorder
($Z t_c/U=0.2385$), this result confirms previous findings
showing that disorder enlarges the superfluid region. At the
critical point, we then study the dynamic conductivity.
\end{abstract}


\maketitle

\section{Introduction} Since the publication of seminal papers on superfluid-insulator transition by Fisher \etal,
\cite{fisher89, fisher90} the disordered two-dimensional (2D)
Bose-Hubbard (BH) model has attracted much theoretical
attention. The disordered BH model with Cooper pairs acting as
charge-2$e$ bosons has been argued to describe the
superconductor-insulator transition in thin amorphous films.
\cite{fisher90,haviland89,jaeger89,cha91} Recently, with the
development of experimental techniques for constructing the BH
model by confining cold alkali atoms in optical lattices,
superfluid (SF) to Mott insulator (MI) transitions have been
observed in clean optical lattices\cite{greiner02}. Further
experiments have introduced disorder in the optical lattices by
speckle fields to investigate the phase diagram of a disordered
three dimensional optical lattice\cite{demarco09a,demarco09b}.
Although experimental techniques still need to be refined,
there is renewed interest in the disordered BH model from
experiment. From theory, there is a long-standing research
interest in studying the disorder induced phase transitions and
searching for Bose glass (BG) phase.
\cite{scalettar91,krauth91,sorensen92,singh92,wallin94,pai96,mukhopadhyay96,freericks96,
svistunov96,herbut97,kisker97,pazmandi98,herbut98,sen01,lee01,cluster,directed,weichman08a,Hitchcock06,wu08,bissbort09,pollet09,kruger09}
In particular, recent work has focused on the transition in 3
dimensions~\cite{Hitchcock06,Sansone07,Gurarie09,kruger09} and
the validity of the relation $z=d$ established by Fisher et
al.~\cite{fisher89} has been questioned both in
numerical~\cite{makivic93,Priyadarshee06} and theoretical
studies~\cite{weichman07}. Here we shall focus exclusively on
the disordered two-dimensional (2D) model where a number of
outstanding questions remain.

\begin{figure}[th]
\begin{center}
\includegraphics[clip,width=\columnwidth]{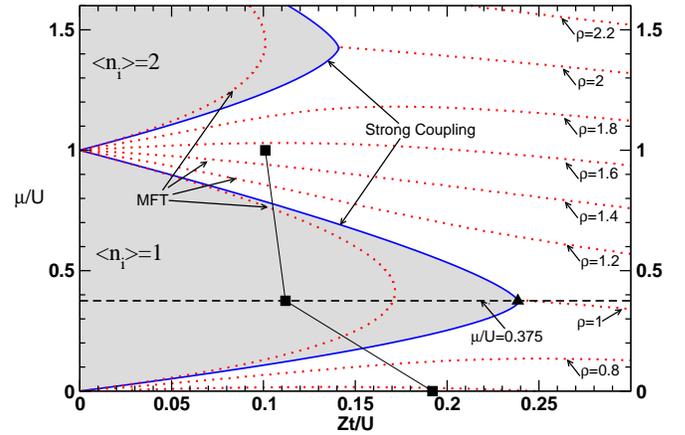}
\caption{ (Color online) The phase diagram of the
$2D$ Bose-Hubbard model. Shaded areas are the Mott-insulating
phases for zero disorder as determined from strong coupling expansions
in Ref.~\onlinecite{Elstner99,Niemeyer99}.
The mean field phase boundaries and constant density profiles for zero
disorder are shown as red dotted lines.
The dashed line indicates the constant chemical potential $\mu/U=0.375$. 
The solid triangle indicates the location of the
transition to the Mott phase in the
  {\it absence} of disorder as determined by SSE simulations
along the dashed line from Ref~\onlinecite{smakov05}. The three solid squares from bottom up are 
for the locations of superfluid to Bose glass transitions in the 
  {\it presence} of disorder at 
$\mu/U=0, 0.375, 1$, respectively, as  determined from SSE simulations in the present work.} \label{phasediag}
\end{center}
\end{figure}

The disordered 2D BH Hamiltonian is given by
\begin{equation}
H=-t\sum_{i,\delta} (a_{i+\delta}^{\dagger}a_i+{\rm H.c.})+\frac{U}{2}\sum_in_i(n_i-1)+\sum_i(\epsilon_i-\mu)n_i,
\label{siterep}
\end{equation}
where $\delta={\bf \hat{x},\hat{y}}$, $t$ is the nearest neighbor hopping
amplitude, $a_i^{\dagger} (a_i)$ is a boson creation
(annihilation) operator, H.c. means Hermitian conjugate,
$n_i=a_i^{\dagger}a_i$ is the number operator, $U$ is on-site
interaction, $\mu$ is chemical potential and $\epsilon_i$ is
uniformly distributed in the interval $[-\Delta, \Delta]$, with
$\Delta$ controlling the disorder strength. Hereafter, we shall
explicitly give energy values for both $t$ and $U$ in each calculation to 
facilitate comparisons with calculations in the literature, since 
Hamiltonian definitions in each paper may be different. 
The phase diagram for this model in the absence of
disorder as determined from strong coupling expansions (from Ref.~\onlinecite{Elstner99,Niemeyer99})
is shown in Fig.~\ref{phasediag} with Mott insulating
lobes with fixed on-site particle number extending into the
superfluid phase. The mean field phase diagram~\cite{muquarter} is shown
as dotted lines in Fig.~\ref{phasediag} and the importance of fluctuations
is clearly evident from the discrepancy between the mean field and strong coupling results
at the tip of the envelopes.
Despite its relative simplicity, a detailed
understanding of this model with disorder has proven
surprisingly difficult in particular for numerical work.

Existing studies using various methods address different
aspects of the disordered BH model, and often arrive at
contradicting conclusions. It is therefore most useful to
revisit this problem using current high performance numerical
techniques.  In the presence of disorder it is
known~\cite{fisher89} that a BG phase appears in addition to
the SF and MI phase present without disorder.  The question of
whether a transition directly from the SF to the MI without an
intermediate BG phase is possible in the presence of disorder
arises. However, this question now seems settled with a proof
that there is always an intermediate BG phase.~\cite{pollet09}
In the simulations we report here, we are always in the strong-disorder 
regime ($\Delta=U$) and we focus on the SF-BG
transition since we expect the MI phases to be strongly suppressed at strong
disorder.

A model closely related to the Bose-Hubbard model is the ($N=2$)
quantum rotors model: $[\cos(\theta_{\bf r}),\sin(\theta_{\bf
r})]$, believed to be in the same universality class as
Eq.~(\ref{siterep}). This model describes a wide range of phase
transitions dominated by phase-fluctuations:
\begin{equation}
H_{\text{qr}}=
\frac{U}{2}\sum_{\bf r}
\left( \frac{1}{i}\frac{\partial}{\partial
\theta_{\bf r}} \right)^2
+i\sum_{\bf r} \mu
\frac{\partial}{\partial \theta_{\bf r}}
-t\sum_{\langle {\bf r},{\bf r'}\rangle }
\cos(\theta_{\bf r}-\theta_{\bf r'}).
\label{eq:hqr}
\end{equation}
Here, $t$ is the renormalized hopping strength and
$\frac{1}{i}\frac{\partial}{\partial\theta_{\bf r}} =L_{\bf r}$
is the angular momentum of the quantum rotor. The angular
momentum can be thought of as describing the deviation of the
particle number from its mean, $L_{\bf r}\simeq n_{\bf r}-n_0$.
This model can be obtained from Eq.~(\ref{siterep}) if
amplitude fluctuations are integrated out when compared to
Eq.~(\ref{siterep}). It is implicitly assumed that only
phase-fluctuations are important at the quantum critical point (QCP),
while amplitude fluctuations are neglected. The model
Eq.~(\ref{eq:hqr}) can be simulated very efficiently using the
Villain (link-current) representation, and a specialized
directed geometrical worm algorithm~\cite{cluster,directed} has
been developed for this purpose. This technique can be applied
to both the clean and disordered model and high-precision
results can be obtained. However, when compared to results obtained by
direct simulations of Eq.~(\ref{siterep}) by
Batrouni~\etal~\cite{batrouni93} as well as from simulations of
Eq.~(\ref{siterep}) in the hard-core limit by
Makivi\'c~\etal~\cite{makivic93}, discrepancies appeared in
particular for the universal features of the conductivity at
the critical point, as we discuss below. Here we shall show
that a likely explanation for these discrepancies is an
inaccurate determination of the QCP in the
direct simulations of Eq.~(\ref{siterep}). Furthermore, it is
important, from our point of view, to provide accurate quantum
critical parameter values (e.g. $t_c$ and $U_c$ for the
Hamiltonian defined in Eq. (\ref{siterep})) using recently
improved simulation techniques.

Our calculations proceed under two different conditions: fixed
particle number (canonical) in order to follow the calculations
of Ref.~\onlinecite{batrouni93} or fixed chemical potential
(grand canonical). This corresponds to taking two different
routes in probing QCP in the BH model phase diagram illustrated
by the dotted and dashed lines in Fig.~\ref{phasediag}.
Previous calculations on the {\it clean} system by \v{S}makov
\etal \cite{smakov05} using a quantum Monte Carlo (QMC) method
have located the QCP at $Zt_c/U=0.2385$ (solid triangle in Fig.
\ref{phasediag}) at $\mu/U=0.375$  (equivalent to $\rho=1$), where $Z=4$ is
the coordination number for the square lattice. As evident from Fig.~\ref{phasediag}
this is in excellent agreement with the strong-coupling result from Ref.~\onlinecite{Niemeyer99,Elstner99}.
Here we follow
a similar approach to determine the QCP but with an added on-site {\it disordered}
chemical potentials $\epsilon_i\in [-\Delta,\Delta]$, with
$\Delta=U$ in the region of strong disorder. 

\begin{figure}[th]
\begin{center}
\includegraphics[clip,width=\columnwidth]{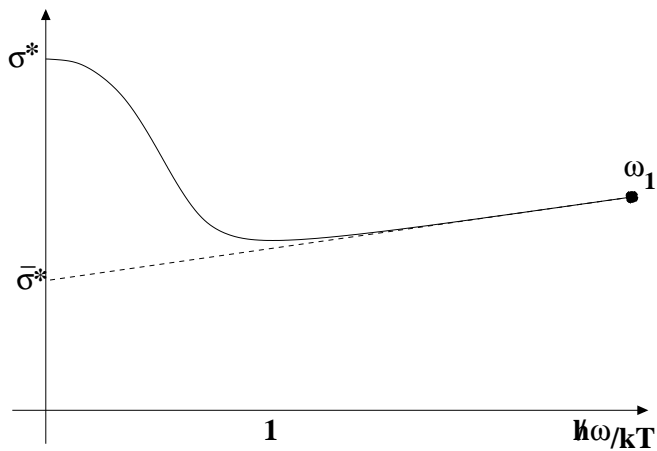}
\caption{A sketch of the conductivity with the two regimes,
$\hbar\omega/(k_BT)\gg 1$ and $\hbar\omega/(k_BT)\ll 1$ clearly
apparent. Also shown is the difference between the universal dc
conductivity $\sigma^*$ and the conductivity $\bar\sigma^*$
obtained from extrapolating the high-frequency numerical data.
An approximate position of the first non-zero Matsubara
frequency is also shown.}\label{sigma}
\end{center}
\end{figure}

At the critical point in a 2D system, the dc conductivity was
predicted \cite{fisher90} to have a universal value,
$\sigma^*$, close to the conductivity ``quantum''
$\sigma_Q=e^{\star 2}/h$, with $e^\star$ the charge of the
bosons. However, the exact universal value has yet to be
determined. Experiments\cite{haviland89,jaeger89} suggest the
dc conductivity value to be a little larger than $\sigma_Q$, but
there are concerns that the experimental temperature (typically
$>0.5$ K) is not low enough. Using the quantum rotor model,
Eq.~(\ref{eq:hqr}), S{\o}rensen \etal \cite{sorensen92} showed
that the universal conductivity value is given by
$\bar\sigma^*=(0.14\pm 0.01)\sigma_Q$. Previously, exact
diagonalization on a hard-core BH model by Runge \cite{runge92}
gave a value of $\bar\sigma^*=(0.15\pm 0.01)\sigma_Q$ in good
agreement. Later, by doing world-line quantum Monte Carlo
(WLQMC) simulations directly on a disordered 2D BH model,
Batrouni \etal \cite{batrouni93} found
$\bar\sigma^*=(0.45\pm0.07)\sigma_Q$ and Makivi\'c
\etal~\cite{makivic93} found $\bar\sigma^*=(1.2\pm
0.2)\sigma_Q$ using a hard-core BH model. These calculations
differ from each other and the experimental value.  It was
later pointed out \cite{damle97} that in $d$ spatial
dimensions, the dynamic conductivity $\sigma(\omega)$ obeys the
following scaling relation near the QCP
\begin{equation}
\sigma(\omega)=2\pi\sigma_Q(\frac{k_BT}{\hbar
c})^{(d-2)/z}\Sigma(\frac{\hbar\omega}{k_BT}),
\label{dynamicsigma}
\end{equation}
which might in part explain the difference between the
experimental value and numerical calculations since the
$\hbar\omega/k_BT\rightarrow 0$ is achieved in the experiments
while in the imaginary-time QMC simulations with Matsubara
frequency $\omega_k$ one always has $\hbar\omega_k/k_BT=2\pi k
>1$. The regimes $\hbar\omega/(k_BT) \ll 1$ and
$\hbar\omega/(k_BT) \gg 1$ are dominated by different transport
mechanisms, hydrodynamic-collision dominated and collisionless
phase-coherent, respectively. Hence, there is little reason to
believe that a simple extrapolation using only
$\hbar\omega/k_BT \gg 1$ can correctly determine the observed
experimental dc conductivity. If a careful extrapolation first
to $L\to\infty$ and then $T\to0$ is performed it is possible to
gain some information about this limit.~\cite{smakov05}
Nevertheless, the entire function $\Sigma$ in
Eq.~(\ref{dynamicsigma}) is universal and the above mentioned
numerical results should still agree on this universal function
and the extrapolations should yield the same number. However,
that extrapolated number which we shall call $\bar\sigma^*$
might not be closely related to the dc conductivity but will
instead correspond to a higher frequency part of $\Sigma$. A
sketch of the expected behavior is shown in Fig.~\ref{sigma}.
Our results here show that if a careful determination of the
QCP is performed, Eq.~(\ref{siterep}) and
Eq.~(\ref{eq:hqr}) yield the same value for $\bar\sigma^*$.

If the conductivity for the BH model at the critical point is a 
scaling function of $\hbar\omega/k_BT$, one would expect plots of 
$\sigma(\omega)$ versus $\omega$ to show {\it deviations} from scaling 
at the critical point even for finite systems. A primary goal of this 
paper is to see if such deviations are observable for the two-dimensional 
disordered BH model if the QCP is determined carefully. 
We believe that the results we present here show clear indications of such 
deviations from scaling with $\omega$.

Not only the conductivity value and dynamic conductivity
scaling differ in historical studies, but also the QCP has
diverse estimates. For example, Zhang \etal \cite{swzhang95}
used both ground-state and finite-temperature QMC simulations
to locate the 
QCP of a hard-core BH
model, which seems to agree with earlier work of Krauth,
Trivedi, and Ceperley. \cite{krauth91} These two simulations
are, however, quite different from what Batrouni \etal found
\cite{batrouni93}. It is noteworthy that these QMC calculations
have used only about 100 disorder samples; we show below that
this is not sufficient for a precise determination of the
critical point.

Recently, QMC algorithms for the BH model have undergone a big
improvement in efficiency with the development of stochastic
series expansion (SSE) algorithm \cite{sandvik92} and the
directed loop-update technique. \cite{sandvik99, sandvik02}
This algorithm does not have the imaginary-time discretization
error inherent in the traditional WLQMC method. The loop
updates are especially important in the quantum critical region
where the long correlation time substantially increases the
errors of the WLQMC simulations. It is now feasible to check
the convergence of measured quantities with the disorder
averaging by increasing the number of disorder samples
dramatically. This has been found by Wallin \etal
\cite{wallin94} and Hitchcock \etal\cite{Hitchcock06} to be
crucial in order to obtain reliable results.

In this paper, we first discuss the SSE algorithm  as it is
applied to the BH model.  In particular, we discuss the
measurement of superfluid density and dynamic conductivity.
Then, tests on the equilibration process and autocorrelation
functions of the simulations are described to validate the
estimates of the superfluid density needed to locate the QCP.
We then show the superfluid density scaling figures for both
the canonical and grand canonical ensembles, discussing the
connection and differences from previous results. Finally, we
show dynamical conductivity scaling and the extrapolated
universal conductivity values in the high-frequency limit.

\section{Numerical Method and Convergence Tests}

As pointed out by Weichman, \cite{weichman08a} analytic
solutions of the model, Eq.~(\ref{siterep}), based on
perturbation of the non-interacting limit has a fundamental
difficulty because in the absence of repulsive interaction $U$
or the Pauli exclusion for fermions, the presence of any
disorder no matter how weak will condense a macroscopic number
of particles into the lowest localized free particle eigenstate
of the random potential. This is not a meaningful state to do
perturbation on. When the interaction $U$ is non-zero, a
complicated competition between the interaction and disorder
potential makes it impossible to do analytic calculations. In
this paper, we resort to QMC simulations of the model with the
SSE algorithm,~\cite{sandvik92,sandvik99,sandvik02}, since it is able to
treat any interaction strength and disorder realization. The
method is briefly discussed below.

In the SSE formalism, the above site representation of Hamiltonian
Eq.~(\ref{siterep}) needs to be written in a bond
representation.
\begin{equation}
H=-\sum_{b=1}^{N_b}(H_{1b}+H_{2b})
\end{equation}
where the $b=\langle ij\rangle$ is the bond index, $H_{1b}$ is a diagonal operator, and $H_{2b}$ is an
off-diagonal operator given by the following equations:
\begin{eqnarray}
H_{1,b}&=&C-[\frac{\tilde{U}}{2}n_i(n_i-1)+\frac{\tilde{U}}{2}n_j(n_j-1)+(\tilde{\epsilon}_i-\tilde{\mu})n_i
\nonumber\\
    & &+(\tilde{\epsilon}_j-\tilde{\mu})n_j],\nonumber \\
H_{2,b}&=&t(a_i^{\dagger}a_j+H.c.),
\end{eqnarray}
where $\tilde{A}=A/Z$, with $A$ one of $\mu,\epsilon_i,U$ and
$Z=4$ being the coordination number of each lattice site.
$N_b=ZN/2$ is number of bonds in the system with $N$ lattice
sites. $C$ is a constant chosen to ensure a positive definite
expansion. Since $\tilde{\epsilon}$ is a random number, we need
to use the disorder amplitude $\Delta$ in order to determine an
appropriate value for the constant $C$.

The partition function can then be expanded as
\begin{eqnarray}
Z&=&{\rm Tr}\{e^{-\beta H}\}=\sum_{\alpha}\sum_{n=0}^{\infty}\frac{(-\beta)^n}{n!}\langle\alpha|H^n|\alpha\rangle,\nonumber \\
&=&\sum_{\alpha}\sum_{n=0}^{\infty}\sum_{S_n}\frac{\beta^n}{n!}\langle\alpha|\prod_{i=1}^nH_{a_i,b_i}|\alpha\rangle,
\end{eqnarray}
where $\{|\alpha\rangle\}$ represents a complete state basis and
\begin{equation}
S_n=[a_1,b_1],[a_2,b_2],\ldots,[a_n,b_n]\nonumber
\end{equation}
is an operator-bond sequence, with $a_i\in\{1,2\}$ denoting the
type of operator (1=diagonal, 2=off-diagonal), and
$b_i\in\{1,\ldots,N_b\}$ being the bond index. The expansion is
now truncated at order $M$, and $M-n$ unit operators for the
$n$th order term are inserted:
\begin{equation}
Z=\sum_{\alpha}\sum_{S_M}\frac{\beta^n(M-n)!}{M!}\langle\alpha|\prod_{i=1}^nH_{a_i,b_i}|\alpha\rangle.
\end{equation}
Here $M$ has been made large enough so that the probability for the expansion order exceeding $M$ can be neglected.
The Monte Carlo weight for the configuration $(\alpha, S_M)$ is then given by:
\begin{equation}
w(\alpha, S_M)=\frac{\beta^n(M-n)!}{M!}\langle\alpha|\prod_{i=1}^nH_{a_i,b_i}|\alpha\rangle.
\end{equation}
The resulting configuration space, consisting of state and operator sequences,
is updated with both diagonal and loop updates.

The diagonal update probability with $n\rightarrow n\pm 1$ is given by
\begin{eqnarray}
P(n\rightarrow n+1)&=&\frac{N_b\beta\langle\alpha(p)|H_{1,b}|\alpha(p)\rangle}{M-n},\nonumber\\
P(n\rightarrow n-1)&=&\frac{M-n+1}{N_b\beta\langle\alpha(p)|H_{1,b}|\alpha(p)\rangle}
\end{eqnarray}
where $|\alpha(p)\rangle$ is the propagated state
\begin{equation}
|\alpha(p)\rangle\sim \prod_{i=1}^p H_{a_i,b_i}|\alpha\rangle.
\end{equation}
For the loop update, transition probabilities during the loop
construction are chosen to have bounce-free or bounce minimizing
solutions of the operator vertex equation. \cite{sandvik02} The
configuration weights for operator vertex in the disordered BH
model are, for particle occupation $n_i$ and $n_j$ for nearest
neighbor $\langle ij\rangle$, given by the following different
cases:
\begin{eqnarray}
& &w(n_i,n_j; n_i+1,n_j-1)=t_{ij}\sqrt{(n_i+1)n_j}(1-\delta_{n_i,n_{\rm max}})\nonumber\\
& &w(n_i,n_j; n_i-1,n_j+1)=t_{ij}\sqrt{n_i(n_j+1)}(1-\delta_{n_j,n_{\rm max}})\nonumber\\
& &w(n_i,n_j; n_i,n_j)=C-[\frac{\tilde{U}}{2}n_i(n_i-1)+\frac{\tilde{U}}{2}n_j(n_j-1)\nonumber\\
  & &+(\tilde{\epsilon}_i-\tilde{\mu})n_i+(\tilde{\epsilon}_j-\tilde{\mu})n_j].
\end{eqnarray}
Here $n_{\rm max}$ is the maximum number of bosons that can
occupy the same site. $n_{\rm max}$ is usually assigned a large
enough value (compared with the average density of the system)
so that it can allow particle number fluctuations while at the
same time will never actually be exceeded during the QMC
simulations. From the operator vertex weight expressions, we
can see that for the disordered BH model, all these weights
need to be calculated ``on the fly''. Note that we have not
explored the possibility of tabulating all the operator vertex
weights, which could speedup the calculation. In the
tabulation, one would need to set up $ZN/2$ probability tables
(each table corresponds to one bond); the number of elements in
the table is determined by $n_{\rm max}$, e.g. for $n_{\rm
max}=4$, each table has 3392 elements.

The total energy of the system is related to the expansion order of the operator sequence
\begin{equation}
E=-\frac{\langle n\rangle}{\beta},
\end{equation}
where $n$ is number of non-unit operators in the operator
sequence as discussed above.

The physical quantity that indicates the superfluid-insulator
transition is the normalized superfluid density $\rho_s$,
\cite{ceperley87} computed as the average square winding
numbers $\langle W^2\rangle$
\begin{equation}
\rho_s=\frac{\langle W^2\rangle}{2t\beta\rho},
\end{equation}
where $W^2=W_x^2+W_y^2$, $\beta$ is inverse temperature, and $\rho$ is the average 
number of particles per lattice site.

To locate QCP, we will use the normalized superfluid density finite-size scaling 
relation \cite{fisher84} 
\begin{equation}
\rho_s=L^{\alpha}f(aL^{1/\nu}\delta,\beta L^{-z}),\label{eq:rhoscal}
\end{equation} 
where $L$ is the linear dimension of the square lattice, $d=2$
is lattice dimension, $\alpha=2-d-z$, $\delta$
measures the distance to the critical point (e.g., to determine $U_c$, we 
have $\delta=(U-U_c)/U_c$, and similarly for $t_c$), $z$ is dynamical
exponent, which is predicted \cite{fisher90} to be $z=2$, $a$
is a non-universal metric number, and the function $f$ is
universal. Hence, we have $\alpha=-2$, and if we keep $\beta
L^{-z}$ fixed, and plot $L^2\rho_s$ versus $Zt/U$ for different
lattice sizes, all the curves will intersect at the critical
value of $U_c$. Note that we assume the validity of the
relation $z=d$ which has been brought into question
recently.~\cite{weichman07}

The conductivity of the BH model as a function of Matsubura
frequency $\omega_k=2\pi k/(\hbar\beta)$ can be calculated from
the linear response relations (we illustrate using the $x$
direction; similar response formulas apply for the $y$
direction) \cite{scalapino92, scalapino93, mahan90}
\begin{equation}
\sigma(i\omega_k)=2\pi\sigma_Q\frac{\langle k_x\rangle-\Lambda_{xx}(i\omega_k)}{\omega_k},
\end{equation}
where $\langle k_x\rangle$ is kinetic energy per link along the $x$ direction, and $\Lambda_{xx}(i\omega_k)$ is the
Fourier transform of the imaginary-time current-current correlation function $\Lambda_{xx}(\tau)$
\begin{equation}
\Lambda_{xx}(i\omega_k)=\frac{1}{N}\int_0^{\beta}d\tau e^{i\omega_k\tau}\Lambda_{xx}(\tau). \label{cccf}
\end{equation}
Here $N$ is total number of lattice sites, and
$\Lambda_{xx}(\tau)=\langle j_x(\tau)j_x(0)\rangle$, with the
paramagnetic current being given by $j_x(0)=it\sum_{\bf
r}(a_{{\bf r}+x}^{\dagger}a_{\bf r}-a_{\bf r}^{\dagger}a_{{\bf
r}+x})$, and $j_x(\tau)$ being the Heisenberg representation of
$j_x(0)$. Note that the smallest Matsubara frequency $\omega_1$
corresponds to $\hbar\omega_1/(k_BT)=2\pi$ significantly larger
than one. Hence, as pointed out by Damle \etal~\cite{damle97}
this type of imaginary-time QMC calculation will necessarily be
in the the collisionless phase coherent regime with
$\hbar\omega/(k_B T)\gg 1$. In the following we set $\hbar=1$.

To calculate the imaginary-time current-current correlation
function $\Lambda_{xx}(\tau)$, we follow the discussion of Ref.
\onlinecite{smakov05} and divide $\Lambda_{xx}(\tau)$ into four
components (which may be thought of as combinations of current
and anti-current terms) by
\begin{equation}
\Lambda_{xx}(\tau)=\sum_{\gamma,\nu=\pm}\Lambda_{xx}^{\gamma\nu}(\tau),
\end{equation}
and
\begin{eqnarray}
\Lambda_{xx}^{\gamma\nu}(\tau)&=&\sum_{\bf r}\langle K_x^{\gamma}({\bf r},\tau)K_x^{\nu}({\bf 0},0)\rangle,\\
K_x^{+}({\bf r},\tau)&=&t a_{{\bf r}+x}^{\dagger}(\tau)a_{{\bf r}}(\tau),\\
K_x^{-}({\bf r},\tau)&=&-t a_{{\bf r}}^{\dagger}(\tau)a_{{\bf r}+x}(\tau).
\end{eqnarray}
Finally the imaginary-time current-current correlation function can be calculated with a binomial summation in QMC
simulations \cite{sandvik92, smakov05}
\begin{eqnarray}
\Lambda_{xx}(i\omega_k)&=&\sum_{\gamma,\nu=\pm}\Lambda_{xx}^{\gamma\nu}(\tau),\nonumber\\
&=&\sum_{{\bf r};\gamma,\nu=\pm}\langle\frac{1}{\beta}\sum_{m=0}^{n-2}a_{mn}(i\omega_k)N(\nu,\gamma;m)\rangle,
\end{eqnarray}
where $a_{mn}(i\omega_k)$ is the degenerate hypergeometric
(Kummer) function,\cite{abramowitz} i.e.,
$a_{mn}(i\omega_k)=$$_1F_1(m+1,n; i\beta\omega_k)$,
$N(\nu,\gamma;m)$ is the number of times the operators
$K_x^{\gamma}$ and $K_x^{\nu}$ appear in the SSE operator
sequence separated by $m$ operators, and $n$ is the expansion
order. The virtue of the above formulation is that the
imaginary-time integral in Eq. (\ref{cccf}) is performed
analytically, hence eliminating the discretization error of the
imaginary-time integral that is present in the conventional
WLQMC. \cite{batrouni93}

\subsubsection{Convergence Tests}
\begin{figure}[th]
\begin{center}
\includegraphics[clip,width=\columnwidth]{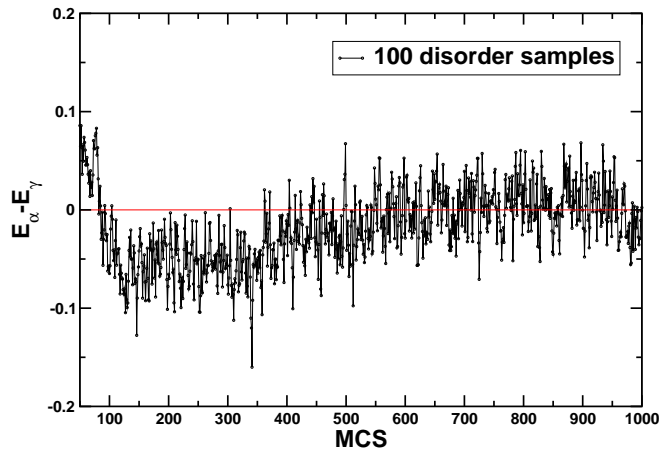}
\caption{The energy difference between two replicas
$E_{\alpha}-E_{\gamma}$ as a function of MCS averaged over 100
disorder samples. Results are shown for $L=16$, $t=0.5$, $U=11$, and
$\beta=12.8$. We start recording the energy difference after 50
MCS.} \label{replicas}
\end{center}
\end{figure}

\begin{figure}[th]
\begin{center}
\includegraphics[clip,width=\columnwidth]{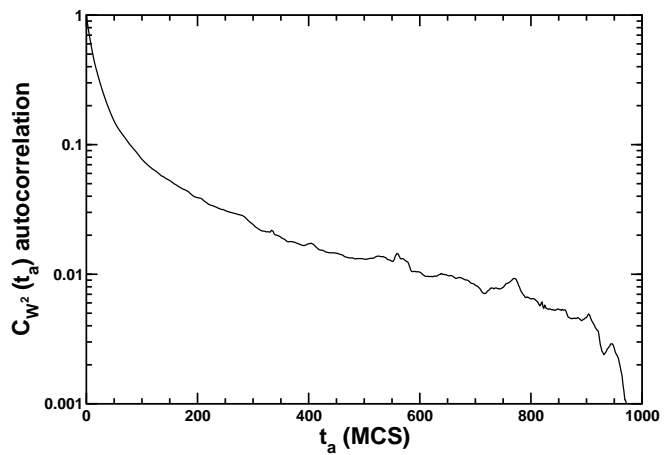}
\caption{The autocorrelation function, $C_{W^2}(t_a)$, for $W^2$
averaged over 100 disorder samples for $L=16$, $t=0.5$, $U=11$, and
$\beta=12.8$. Note that the $y$-axis is on a $\log$ scale.}
\label{autocorr}
\end{center}
\end{figure}

Numerically we perform SSE simulations for lattice sizes $L=6,
8, 10, 12, 16$. According to scaling relation Eq. (\ref{eq:rhoscal}), 
the corresponding inverse temperatures for
these lattices are $\beta=1.8, 3.2, 5.0, 7.2, 12.8$, leading to
a constant $\beta L^{-2}=0.05$. Our strategy is to ensure that
each disorder realization has been equilibrated, resulting in a
few statistically independent measurements for each disorder
realization, with the error-bars then obtained from the
disorder averaging where each realization can be considered
statistically independent. We have performed several tests to
ensure that this equilibration is attained and that the number
of disorder realizations are sufficient~\cite{Hitchcock06}.

We begin by considering the equilibration of the energy between
two different replicas with the {\it same} disorder
realization. We define one Monte Carlo sweep (MCS) as 1
diagonal update, which includes a sweep through all diagonal
operators in the SSE expansion, followed by 10 loop updates
(the number of loop updates included in 1 MCS is arbitrary, but
it is usually determined by the ratio of the SSE expansion
order to the average loop length). In Fig. \ref{replicas}, we
show the equilibration of the energy difference between two
energy replicas as a function of MCS for $L=16$, $t=0.5$, $U=11$
averaged over 100 disorder realizations. As mentioned, replica
simulation means that for each disorder realization, we start
two parallel SSE simulations $\alpha$ and $\gamma$ with vastly
different initial configurations but with the {\it same}
disorder realization, and monitor the evolution of
$E_{\alpha}-E_{\gamma}$. We see that after about 500 MCS,
$E_{\alpha}-E_{\gamma}$ fluctuates around zero, showing
equilibration.

\begin{figure}[th]
\begin{center}
\includegraphics[clip,width=\columnwidth]{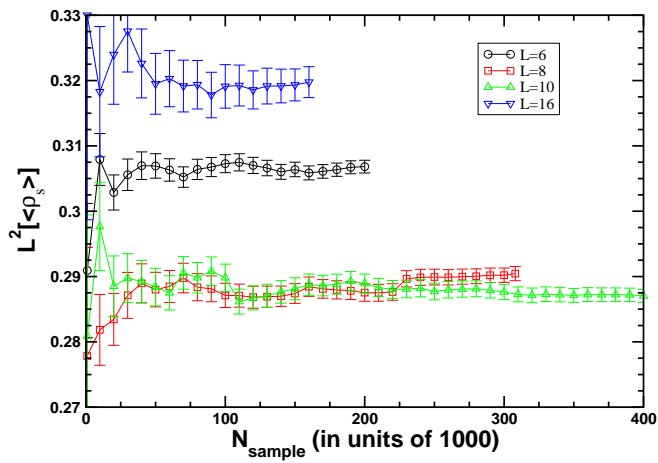}
\caption{(Color online) Convergence of the superfluid density
for $t=0.5$ and $U=11$ as a function of the number of disorder samples
$N_{\rm sample}$ for various lattice sizes, $L=6,8,10,$ and
16.} \label{disorder_converge}
\end{center}
\end{figure}

We then proceed check the convergence of the winding number
$W^2$ by calculating its autocorrelation function,
\begin{equation}
C_{W^2}(t_a)=\frac{\langle W^2(0)W^2(t_a)\rangle -\langle W^2\rangle^2}{\langle W^4\rangle-\langle W^2\rangle^2}.
\end{equation}
Our results for $C_{W^2}(t_a)$ versus $t_a$ (MCS) are shown in Fig.
\ref{autocorr} on a  $\log$ scale averaged over 100 disorder
samples. Since we expect the leading term in $C_{W^2}(t_a)$ to
scale as $e^{-t_a/\tau}$, we estimate the
autocorrelation time to be around $\tau\sim 40$ MCS.

Based on the above tests we set the following parameters for
the SSE simulations and measurements: for each disorder
realization we perform 1000 MCS to warm up the system, which is
followed by 1000 MCS measurements with each measurement being
separated by 1 MCS. While this might seem inadequate to obtain
small error bars for a single disorder realization, this
approach is in fact optimal since each disorder realization is
statistically independent; reliable error-bars can then be
obtained from the disorder averaging. We have done tests for
the worst case, i.e., longest warm-up MCS and $W^2$
autocorrelation time for $L=16$, $t=0.5$, $U=11$, and $\beta=12.8$, so
we expect the parameters to be sufficient for the other lattice
settings.

Finally, we focus on the disorder averaging which we denote by
$[\cdot]$ to distinguish it from thermal averages $<\cdot>$.
Often one finds that the estimate of a quantity one wishes to
calculate obeys a rather broad distribution with a substantial
tail~\cite{Hitchcock06} with a difference between the average
and the most likely values. This will be reflected in an
underestimation of the error-bar calculated from the standard
deviation and a slow ``movement" of the average as the number
of disorder realizations is increased with the error-bar
essentially constant with respect to the number of averages. We
check our convergence here by focusing on the behavior of the
superfluid density as we increase the number of disorder
samples $N_{\rm sample}$. See Fig. \ref{disorder_converge}. For
each curve in the figure, the first point is for 1,000 disorder
realizations, the second for 10,000 disorder realizations, and
for subsequent points, the increment is 10,000 disorder
samples. It is important to note that 1,000 disorder
realizations is already {\it an order of magnitude more} than
the previous calculations,\cite{krauth91,batrouni93,swzhang95}
where the number of disorder realizations was taken to be
around 100. Here, we find that 1,000 disorder samples are still
far from convergence. The idea of self-averaging does not seem
to apply here, either. For the remainder of our results we have
used in excess of 100,000 disorder realizations to ensure
reliable results.

\section{Results}

\subsection{Fixed particle number, $\rho=0.5$, $1$}

As described, the SSE is essentially a grand canonical
simulation algorithm and suitable for fixed chemical potential
QMC simulations. However, in order to be as close as possible
to previous calculations by Batrouni \etal,~\cite{batrouni93}
we also want to do simulations with fixed particle number for
each disorder realization. It is possible to forbid particle
number fluctuation during the loop construction,
\cite{rombouts06a, rombouts06b} but we will adopt a simpler
approach. We first do some rough estimates of chemical
potentials to yield total particle number close to the required
value for a set of disorder realizations. We then use the
average chemical potential for all disorder realizations in the
construction of the probability table during the loop update,
and reject any closed loops that change the total particle
number. In practice, we find that with the roughly estimated
chemical potential the loss of efficiency caused by additional
rejection is insignificant, typically $5\%$ more rejection than
the grand canonical simulations.
\begin{figure}[th]
\begin{center}
\includegraphics[clip,width=\columnwidth]{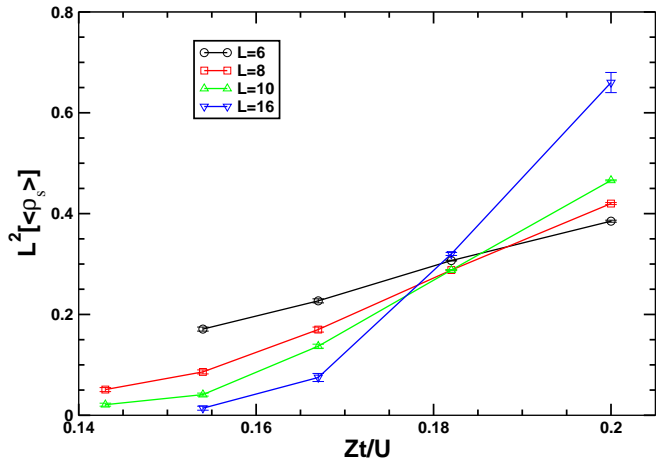}
\caption{(Color online) Superfluid density scaling from 
canonical simulations, i.e., fixed particle number with density $\rho=0.5$, for 
$\Delta=12$ and $t=0.5$, for various lattice sizes, $L=6,8,10,$ and 16.} \label{rhosvsu}
\end{center}
\end{figure}

We now turn to our results using SSE simulations in the canonical ensemble with
$n_{\rm max}=4$, fixed particle density $\rho=0.5$, and fixed disorder strength
$\Delta=12$ and hopping integral $t=0.5$, which are the same as in Ref. \onlinecite{batrouni93}.
Note that canonical systems with fixed particle density of $\rho=0.5$ are very close  
to the grand canonical systems with $\mu/U=0$.
A series of $U$ values will be used in the simulations to search for the
critical point $U_c$ based on finite-size scaling relation for the normalized superfluid
density Eq. (\ref{eq:rhoscal}). 

To locate the QCP $U_c$ or equivalently $Zt/U_c$, we plot in Fig. \ref{rhosvsu} the
superfluid density $L^2\rho_s$ as a function of $Zt/U$ (fixing $t=0.5$ 
and using a $U$ grid of $U=11, 12, 13,$ and 14) for various lattice
sizes $L=6,8,10,$ and 16. All 4 curves intersect very close to
the point $Zt/U_c=0.182$. The crossing of the curves is not as
``perfect" as for the clean system where larger system sizes
can be simulated and it would appear that corrections to
scaling for the smallest size $L=6$ might be sizable. However,
for disordered systems the results in Fig.~\ref{rhosvsu} is at
par with the best one can obtain. We also note that the fact
that we do obtain a crossing of the curves at a single point is
an indirect validation of the scaling relation
Eq.(\ref{eq:rhoscal}) with $z=d$. Based on the data shown in
Fig.~\ref{rhosvsu}, we estimate the critical point to be
$Zt/U_c=0.182\pm 0.003$ or $U_c=11.0\pm 0.5$. To achieve a more accurate result will
remain a challenge for some time. Since canonical simulations with $\rho=0.5$ correspond 
closely to the grand canonical simulations with $\mu/U=0$, we can compare the QCP obtained 
in these two simulations, and find that this canonical QCP value for $\rho=0.5$ is 
very close to the grand canonical value of $(Zt/U)_c=0.192$ at 
$\mu/U=0$, which is shown as black square at the bottom of Fig. \ref{phasediag}.

The same QCP in Ref.~\onlinecite{batrouni93} (note factor of $t/2$ 
instead of $t$ in their Hamiltonian) was estimated to
be $(U/t)_c\sim 7$, substantially different from the result we
obtain here \footnote{Due to possible 
differences in the definition of the hopping term it is possible that their
critical point $(U/t)_c=7$ in our units should be $(U/t)_c=14$. G. Batrouni 
private communication.}. However, for $\rho=0.75$
Ref.~\onlinecite{krauth91} (note factor of $t/2$ instead of $t$ 
in their Hamiltonian) finds $(U/t)_c\sim 10$ and in the
hard-core limit Ref.~\onlinecite{swzhang95} (note factor of $U$ in stead of 
$U/t$ in their Hamiltonian) finds
$(U/t)_c=9.9\pm0.4$ with $\rho=0.5$ in reasonable agreement with our
results. We note that in Ref.~\onlinecite{batrouni93} the
location of the QCP was obtained partly by requiring the
frequency dependent conductivity, $\sigma(\omega)$, to scale
with $\omega$ at the critical point in violation of
Eq.~(\ref{dynamicsigma}). We now turn to a discussion of our
results for the dynamic conductivity.

\begin{figure}[th]
\begin{center}
\includegraphics[clip,width=\columnwidth]{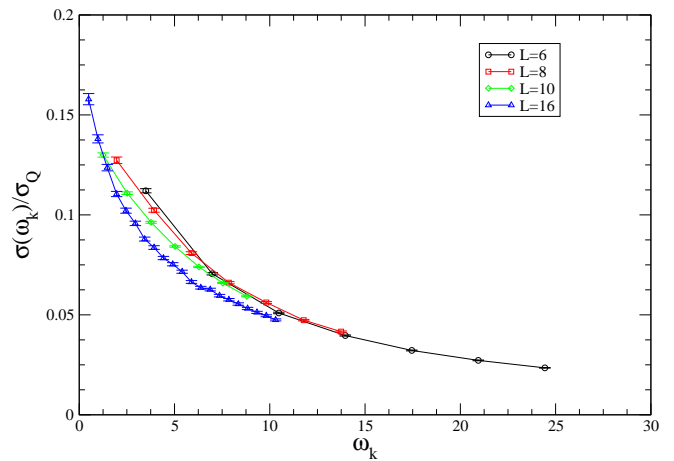}
\caption{(Color online) Dynamic conductivity scaling plot
$\sigma(\omega_k)/\sigma_Q$ vs Matsubara frequency $\omega_k$
for $Zt/U_c=0.182$ and $t=0.5$.} \label{conductU11}
\end{center}
\end{figure}

As discussed above, the dynamic conductivity calculated with
Matsubara frequency is expected to satisfy the scaling relation
Eq. (\ref{dynamicsigma}). The implication of this scaling law
for numerical simulations is that $\sigma(\omega,T,L)$ in the
limit $L\to\infty$ as $T$ approaches 0 should be a function of
a single variable $\hbar\omega/(k_BT)$. Therefore, if the
dynamic conductivity curves calculated from different lattice
sizes are plotted versus $\omega$, one would expect deviations
from a single curve to be visible at the smallest Matsubara
frequencies since finite size dependence there is likely the
smallest and the data should be close to scaling with
$\hbar\omega/(k_BT)$ even without extrapolation to
$L\to\infty$. See also Ref.~\onlinecite{smakov05}. Such behavior is clearly visible in
Fig.~\ref{conductU11}, where SSE results for  the dynamic
conductivity $\sigma(\omega_k)/\sigma_Q$ at the critical point
$Zt/U_c=0.182$ are shown. All curves from 4 different lattice sizes
overlap with each other at high frequency limit but differ at
the low frequency side. A rough estimate of the high frequency
universal conductivity, $\bar\sigma^*$ using the lattice size
$L=16$ gives $\bar\sigma^*\sim 0.17\sigma_Q$.
This is in surprisingly good agreement with simulations of the
quantum rotor model, Eq.~(\ref{eq:hqr}), where one finds
$\bar\sigma^*=(0.14\pm0.01)\sigma_Q$, ~\cite{sorensen92} as
well as with the exact diagonalization result $(0.15\pm
0.01)\sigma_Q$ for a hard-core BH model, \cite{runge92} as one
would expect for a universal quantity.  However, this value is
much less than the value found by Batrouni \etal $(0.47\pm
0.08)\sigma_Q$. \cite{batrouni93} The discrepancy could be due
to an inaccurate estimate of the QCP in
Ref.~\onlinecite{batrouni93} obtained partly by assuming that
$\sigma(\omega)$ should scale with $\omega$ at the quantum
critical point.
\begin{figure}[th]
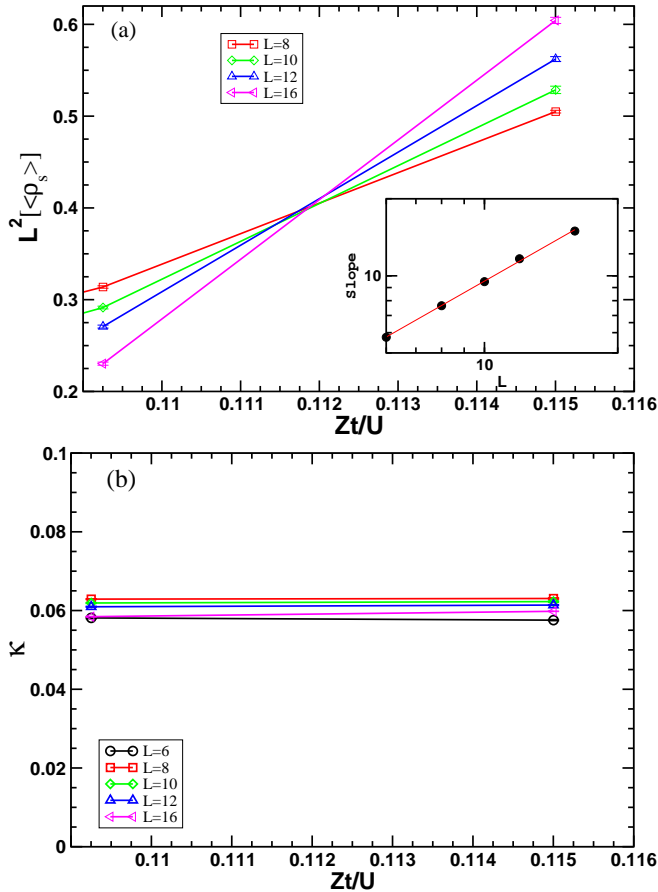

\begin{center}
\begin{tabular}{c}
\includegraphics[clip,width=\columnwidth]{fixed_mu_scaling}\\
\includegraphics[clip,width=\columnwidth]{fixed_mu_kappa}\\
\end{tabular}
\caption{(Color online). (a) Scaling curves of the superfluid
density from grand canonical simulations at $\mu/U=0.375$. The
inset shows the slope of these curves at the QCP on a log-log
scale. (b) The compressibility $\kappa$ in the same region.}
\label{commensurate}
\end{center}
\end{figure}

In addition we have performed canonical simulations with 
$\rho=1$, and determined the QCP at 
[$(Zt/U)_c=0.108(2)$], which is slightly different from the fixed
chemical potential ($\mu/U=0.375$) analysis value of $(Zt/U)_c=0.112(1)$ 
to be discussed in the next section.

\subsection{Grand canonical simulations $\mu/U=0,0.375,1$}
Close to the  commensurate filling of $\rho=1$ corresponding to $\mu/U=0.375$
the QCP has been determined in Ref. \onlinecite{smakov05} for the {\it clean}
BH model (see also solid triangle in Fig. \ref{phasediag}).  Here we are 
interested in further investigating the effect of
disorder on this transition by keeping all the other parameters the same as in
Ref. \onlinecite{smakov05} while setting the disorder potential amplitude
$\Delta=U$.

Note that in generating the disordered onsite potential
$\epsilon_i$ for the system, we make sure that the
average of the disordered onsite potential $\sum_i\epsilon_i/N=0$
for each disordered configuration of the system. This can be
achieved by first generating a disorder configuration and
calculating the average of the disordered potential, then
subtracting this average from the generated potential at each
lattice site.

Our results are shown in Fig. \ref{commensurate}(a) where
$L^2[<\rho_s>]$ is plotted for different system sizes close to
the QCP. The scaling law here is the same as
outlined above for the simulations in the canonical ensemble
and again we note that the relation $z=d$ was assumed. As can
be seen from Fig.~\ref{commensurate}(a) the curves cross very
nearly in a single point and it is easy to estimate the
corresponding QCP points as $(Zt/U)_c=0.112(1)$.  Very
interestingly, when this is compared to the results for the
clean system QCP of $(Zt/U)_c=0.2385$ from Ref.
\onlinecite{smakov05}, it is clear that the introduction of
disorder at this commensurate filling has enhanced the
superfluid region. The QCP with ($\blacksquare$) and without 
($\blacktriangle$) disorder at $\mu/U=0.375$ are shown along the dashed line in 
Fig. \ref{phasediag}. A similar indication for disorder-enhanced
superfluidity has been found by Krauth \etal,~\cite{krauth91}
but the effect is even more noticeable in our results.

Also shown in Fig.~\ref{commensurate}(b) (as an inset) is the
slope of $L^2\rho_s$ at the QCP. From the scaling relation
Eq.~(\ref{eq:rhoscal}) it is clear that this slope should scale
with $L^{1/\nu}$. The solid line in the inset is a fit to the
data yielding $\nu\sim 1$ in agreement with the quantum version
of the Harris criterion $\nu\ge 2/d$.

In Fig.~\ref{commensurate}(b) are shown results for the
compressibility, $\kappa$, through the QCP.
As can be seen, the compressibility remains constant through
the QCP. It is expected that in the presence of disorder the
compressibility will scale as $\kappa\sim\delta^{\nu(d-z)}$.
The fact that our results for $\kappa$ remain constant through
the transition is consistent with the relation $z=d$, although
it does not rule out $z<d$.

\begin{figure}[th]
\begin{center}
\includegraphics[clip,width=\columnwidth]{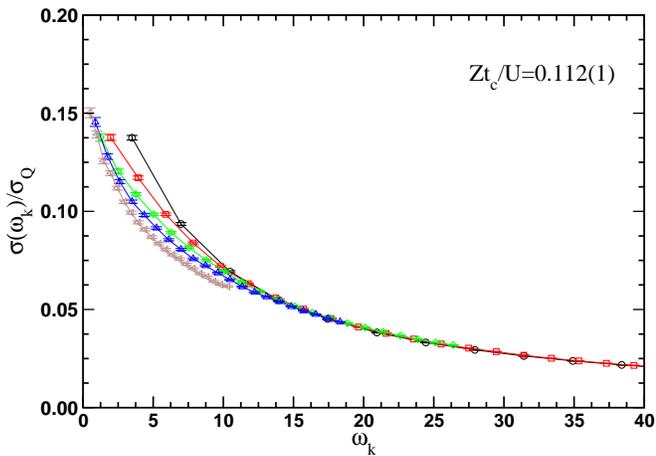}
\caption{(Color online) Dynamic conductivity
$\sigma(\omega_k)/\sigma_Q$ vs Matsubara frequency $\omega_k$
for $Zt_c/U=0.112$ in a disordered grand canonical system at 
$\mu/U=0.375$.}
\label{fixed_mu_sigma}
\end{center}
\end{figure}

At the QCP we also measure the dynamic conductivity with grand
canonical QMC simulations, shown in Fig. \ref{fixed_mu_sigma}.
Like the simulations in the canonical ensemble, we observe that
the data appear to scale with $\omega_k$ at high frequency but
very clearly deviate from this behavior at low frequencies.
Again we emphasize that this is consistent with
$\sigma(\omega)$ scaling with $\hbar\omega/(k_BT)$ rather than
$\omega$. Here we can also attempt to estimate $\bar\sigma^*$
and we find it to be around 0.17$\sigma_Q$, in very good
agreement with the results for the canonical simulations from
the previous section as well as with the simulations of the
quantum rotor model.~\cite{sorensen92} As before, we note that
the universal conductivity, $\bar\sigma^*$ estimated here and
in the previous section corresponds to the conductivity value
in the high frequency limit, not the DC conductivity measured
experimentally.

We have also performed simulations at $\mu/U=1$ where we find
$(Zt/U)_c=0.101$ as well as for $\mu/U=0$ where we find
$(Zt/U)_c=0.192$. These QCPs are also shown as
solid black squares in Fig.~\ref{phasediag}.

\section{Conclusion}
In this paper we have determined the superfluid-insulator
transition QCPs of the 2D BH model with
strong disorder for both commensurate $\rho=1$ and
incommensurate $\rho=0.5$ systems. In connection with previous
calculations in the literature we have identified the
disorder-induced superfluidity at $\rho=1$ in an otherwise Mott
insulating region in the clean limit.  Our main results are
summarized in Fig.~\ref{phasediag} where the dramatic results
of disorder are clearly visible. While in the past, high
precision results have been available for the quantum rotor
model as well as the BH model in the hard-core limit, the
results we present here are direct simulations of the
full BH model with large disorder. Our results clearly indicate
that these three models in the presence of disorder are in the same
universality class.


At the QCP, we also compute the dynamic conductivity as a
function of the Matsubara frequency. The universal conductivity
at high frequency limit is estimated to be around
0.17$\sigma_Q$ in good agreement with previous calculations on
quantum rotor models.~\cite{sorensen92}  Most notably, at the
QCP, the dynamic conductivity shows clear deviations from
scaling with $\omega$, consistent instead with the expected
scaling with $\omega/(k_BT)$. In addition, we checked in
detail the convergence with the number of disorder realizations
finding that a large
number of disorder realizations (usually on the order of
10$^4$) are necessary to have a fully convergent superfluid
density value. This implies that some previous calculations in
the literature with disorder realizations of the order of 10$^2$,
might not be reliable. This has allowed us to determine the QCP
for several values of $\mu/U$ for the disordered Bose-Hubbard
model.

\section{Acknowledgments}
We would like to acknowledge useful correspondence with G. Batrouni, 
R. Scalettar, and J. \v{S}makov. We also thank H. Monien for permission to
use data from Ref.~\onlinecite{Elstner99,Niemeyer99}.
FL was supported by the US Department 
of Energy under award number DE-FG52-06NA26170 when this study was carried out. ESS
is supported by the Natural Sciences and Engineering Research
Council of Canada and the Canadian Foundation for Innovation.
DMC was supported by the DARPA OLE program. Computation time was 
provided by the SHARCNET and NCSA supercomputing facilities.

\bibliography{DBH2dbib}
\end{document}